\documentclass[twocolumn,superscriptaddress,showpacs,preprintnumbers,nofootinbib,amsmath,amssymb,floatfix,aps]{revtex4-1}
\usepackage{dcolumn}
\usepackage{bm}
\usepackage{color}

\usepackage{graphicx}
\usepackage{amsmath}
\usepackage{amssymb}

\def\lsim{\buildrel < \over {_{\sim}}}
\def\gsim{\buildrel > \over {_{\sim}}}

\def\beq{\begin{equation}}
\def\eeq{\end{equation}}
\def\be{\begin{eqnarray}}
\def\ee{\end{eqnarray}}

\begin{document}
\title{Unified description of electron-nucleus scattering\\ within the spectral function
formalism}
\author{Noemi Rocco}
\affiliation{INFN and Department of Physics, ``Sapienza'' University, I-00185 Roma, Italy}
\author{Alessandro Lovato}
\affiliation{Physics Division, Argonne National Laboratory, Argonne, Illinois 60439, USA}
\author{Omar Benhar}
\affiliation{INFN and Department of Physics, ``Sapienza'' University, I-00185 Roma, Italy}
\affiliation{Center for Neutrino Physics, Virginia Tech, Blacksburg, Virginia 24061, USA}

\date{\today}
\begin{abstract}
The formalism based on factorization and nuclear spectral functions has been generalized to treat transition matrix 
elements involving two-nucleon currents, whose contribution to the nuclear electromagnetic response in the transverse 
channel is known to be significant. We report the results of calculations of the inclusive electron-carbon cross section, showing 
that the inclusion of processes involving two-nucleon currents appreciably improves the agreement between theory and data in the 
dip region, between the quasi elastic and $\Delta$-production peaks. The relation to approaches based on the independent particle of the nucleus and 
the implications for the analysis of neutrino-nucleus cross sections are discussed.
\end{abstract}
\maketitle

The nuclear response to electromagnetic interactions is determined by a variety of mechanisms\textemdash reflecting both nuclear and nucleon excitation modes\textemdash whose contributions strongly depend on the energy and momentum transfer, $\omega$ and ${\bf q}$.

In the kinematical region corresponding to
$|{\bf q}| \gsim \pi/d$, with $d$ being the average nucleon-nucleon (NN) distance,  interactions predominantly involve individual nucleons
and, depending on energy transfer, may give rise to different hadronic final states. At $\omega \approx Q^2/2m$, where
$Q^2 = {\bf q}^2 - \omega^2$ and $m$ is the nucleon mass, the dominant mechanism is quasi elastic scattering, in which the nucleon is left in its ground state and no $\pi$-mesons are produced.
  With increasing  $\omega$, the composite nature of the nucleon shows up through the excitation of resonances\textemdash the most prominent
  of which is the $\Delta$, with mass $m_\Delta$= 1232 MeV\textemdash and breakup of the nucleon itself, followed by hadronization of the debris.
The corresponding final states are characterized by the presence of one or more $\pi$-mesons, respectively.

Reaction mechanisms involving two target constituents\textemdash for example the process in which the virtual photon couples to
a meson exchanged between interacting nucleons\textemdash also play an important role in determining the
nuclear response in the transverse channel. They have long been shown to  provide a significant amount of strength in the dip region,
between the quasi elastic and $\Delta$-production peaks~\cite{vanorden}.

The occurrence of two-nucleon components in the nuclear electromagnetic current is dictated by current conservation and the isospin dependence
of nuclear interactions \cite{riska}.  Therefore, a coherent treatment of one- and two-nucleon current contributions, including the effects of interference between the corresponding transition matrix elements, is needed to achieve a complete description of the observed electron-nucleus cross sections.

The {\em ab initio} approach based on nuclear many-body theory and realistic nuclear hamiltonians\textemdash strongly constrained by 
the properties of two- and three-nucleon systems\textemdash provides a fully consistent framework for the calculation of the 
nuclear electromagnetic responses in the regime of low to moderate momentum transfer, typically $|{\bf q}| \lsim$ 500 MeV, in which the nuclear initial and final states can be described within the non relativistic approximation, and the non relativistic reduction of the currents is expected to 
be applicable \cite{GPC,2NC,Carlson:2002,lovato12C, inversion}.

At high momentum transfer, however, neither the nuclear final state nor the current can be treated using the non relativistic formalism, because the former 
involves at least  one nucleon carrying large momentum, $\sim {\bf q}$, while the latter  explicitly depends on ${\bf q}$.  
To circumvent this difficulty,  theoretical calculations of the two-nucleon current contributions to the nuclear cross section have been carried out within somewhat oversimplified models, in which 
relativistic effects are taken into account at the expenses of a realistic description of  nuclear structure and dynamics \cite{alberico,dekker,oset,depace}.

The formalism based on factorization of the nuclear transition matrix elements \cite{PRD,RMP,AHEP} allows to 
combine a fully relativistic description of the electromagnetic interaction with an accurate treatment of nuclear dynamics, in which the effect of 
NN correlations is properly taken into account. This scheme, providing a remarkably accurate description of the available data 
in the kinematical region in which quasi-elastic single-nucleon knock out is the dominant reaction mechanism \cite{uncertainty},
has been recently generalized to include the contributions of  the two-nucleon current \cite{BLR}. 

The analysis of Ref.~\cite{BLR} was restricted  to electromagnetic response of carbon  in the transverse channel, and neglected final state interactions (FSI) between  the struck nucleon and the spectator particles altogether. In this letter, we report the results of calculations of the {\em inclusive} electron-carbon cross section carried out including the contributions of 
one- and  two-body currents, including both elastic and inelastic channels, as well as FSI effect. We emphasize that in our work the relativistic treatment of the two-nucleon current is associated, 
for the first time, with a description 
of the dynamics that goes beyond the independent particle model of the nucleus.

In interacting many-body systems, processes involving one- and two-nucleon currents are inextricably 
related, as they both lead to the appearance of two particle-two hole (2p2h) final states. As a consequence, the corresponding transition amplitudes interfere, and must be treated in a consistent fashion \cite{Carlson:2002}.

Neglecting the contribution of final states involving more than two nucleons in the continuum, the cross section can be written as
\begin{equation}
d\sigma \propto L_{\mu\nu} W^{\mu\nu} = L_{\mu\nu} (W^{\mu\nu}_{1p1h}+ W^{\mu\nu}_{2p2h})\ ,
\label{sigma:split}
\end{equation}
where the label ${\rm n}p{\rm n}h$ refers to n-particle\textendash n-hole final states and the tensor $L_{\mu\nu}$ is completely determined by lepton kinematics. 
The target response tensor $W^{\mu\nu}$,  on the other hand, is written in terms of matrix elements of the nuclear current operator between the target ground state and the 
hadronic final states.

The current entering the definition of the 2p2h component $W^{\mu\nu}_{2p2h}$  can be written in momentum space in the form
 \begin{align}
J^\mu({\bf k_1},{\bf k_2}) & = j^\mu_1({\bf k_1})  \delta({\bf k_2})+  j^\mu_2({\bf k_2})  \delta({\bf k_1})  + j_{12}^\mu({\bf k_1},{\bf k_2})  \ ,
\label{2b:curr}
 \end{align}
 clearly showing how the  total momentum transfer, ${\bf q}~=~{\bf k_1}~+~{\bf k_2}$,  is shared between the two nucleons involved in the electromagnetic interaction, labeled by the indices $1$ and $2$.

\begin{figure}[h!]
\includegraphics[scale=0.425]{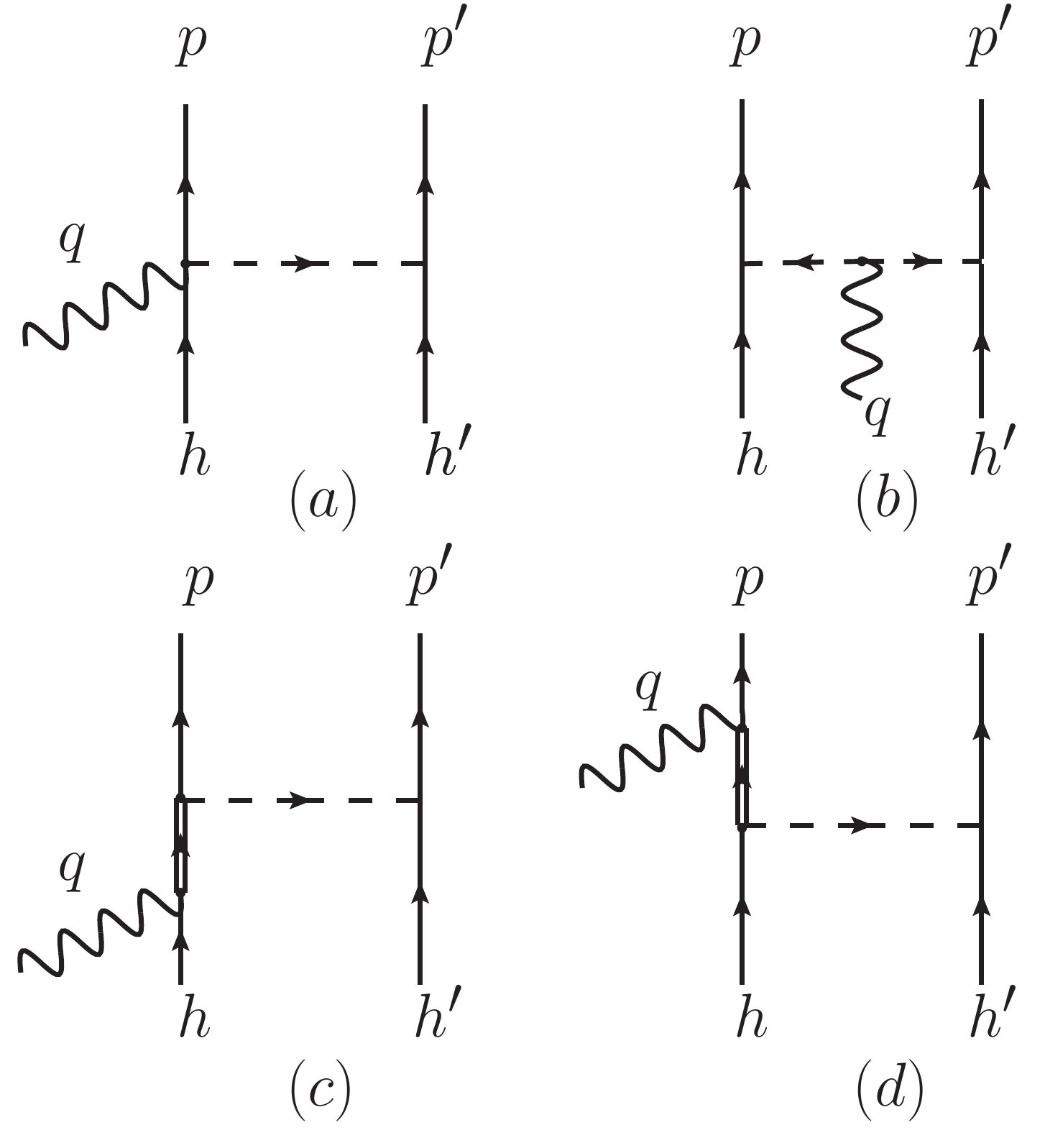}
\vspace*{-.1in}
\caption{Free-space meson exchange current diagrams. The first two correspond to $\pi$-meson exchange: diagram (a) and the one obtained interchanging particle 1 and 2, represent 
the contact or seagull current, while diagram (b) the pion-in-flight current. The diagrams (c)-(d) and  the additional two resulting from the interchange $1\leftrightarrow$ 2, contain an intermediate $\Delta$-isobar excitation.}
\label{mec}
\end{figure}

The Feynman diagrams of Fig.~\ref{mec}, in which ${\bf h}, {\bf h^\prime}$ and ${\bf p},{\bf p^\prime}$ label the momenta of the hole and particle states, respectively,
illustrate the different contributions to $j_{12}^\mu({\bf k_1},{\bf k_2})$.
In this work we have considered two-body currents of two types. The one associated with the exchange of a $\pi$-meson is required by current conservation. Hence, its expression is determined\textemdash at least in principle\textemdash by the structure of the NN potential appearing in in the nuclear hamiltonian.

Diagram (b), featuring  a $\gamma\pi\pi$ vertex, is associated with the ``pion-in-flight'' term, while 
the sum of diagrams (a), involving a $\gamma\pi NN$ vertex, and  that obtained interchanging particles 1 and 2 accounts for  the  ``seagull'', or ``contact'' contribution.
Diagrams (c) and (d), as well as the corresponding two in which particles 1 and 2 are interchanged, are associated with two-body current terms involving a $\Delta$-resonance in the intermediate state. 
Owing to the purely transverse nature of this current, their form is not subject to current-conservation constraints, and is therefore largely model dependent \cite{riska}.
In order to make contact between our results and those obtained by Dekker et al~\cite{dekker} and De Pace et al~\cite{depace}, we have used the fully relativistic expression of the two-body currents reported in their papers, with the same form factors and $\Delta$-width.

The factorization {\em ansatz} amounts to writing the matrix elements describing transitions from the ground state to 2p2h final states
in terms of nuclear amplitudes and matrix elements of the one- and two-body current operators between free-nucleon states. For the one-nucleon current one finds \cite{AHEP,BLR} 
 \begin{align}
\label{fact1b}
\langle 0| j^\mu_1 | {\bf h} {\bf h^\prime} {\bf p} {\bf p^\prime} \rangle= \int d^3 k\  \Phi^{h h^\prime p^\prime}_{k}    \langle{\bf k}|j^\mu_1 |{\bf p}\rangle , 
\end{align}
where the state $|{\bf k}\rangle$ describes a free nucleon carrying momentum ${\bf k}$,  while the overlap between the target ground state and  the 2h1p state of the residual (A-1)-particle system, in which one nucleon is excited to a continuum state outside the Fermi sea, is written in the form
\begin{align}
\label{def:phi2h1p}
\Phi^{h h^\prime p^\prime}_{k} = \langle 0 | \{ | {\bf k} \rangle \otimes | {\bf h} {\bf h}^\prime {\bf p}^\prime  \rangle   \} \ .
\end{align}

Application of the same scheme to the matrix element of the two-nucleon current operator leads to the expression~\cite{BLR}
 \begin{align}
 \langle 0| j^\mu_{12}|  {\bf h} {\bf h^\prime} {\bf p} {\bf p^\prime} \rangle = \int d^3k\ d^3k^\prime\ \Phi^{h h^\prime}_{kk^\prime} \langle {\bf k}{\bf k}^\prime| j^\mu_{12}| {\bf p}{\bf p}^\prime\rangle\ ,\end{align}
with the nuclear amplitude, corresponding to 2h bound states of the (A-2)-nucleon spectator system, given by 
\begin{align}
 \Phi^{h h^\prime}_{k k^\prime}   = \langle 0 | \{ | {\bf k} {\bf k}^\prime \rangle \otimes | {\bf h} {\bf h}^\prime   \rangle   \} \ .
\label{2hov0} 
\end{align}

Using the above results, the 2p2h contribution to the nuclear response tensor can be decomposed according to 
  \begin{align}
 \label{decompose} 
  W^{\mu\nu}_{2p2h}&=  W^{\mu\nu}_{2p2h,11}+ W^{\mu\nu}_{2p2h,22}+ W^{\mu\nu}_{2p2h,12}\ .
 \end{align}
The first term comprises the squared amplitudes involving only the one-nucleon current. Note that the occurrence of
these matrix elements is a genuine correlation effect, not accounted for within the independent particle model.
As a consequence, the calculation of $W^{\mu\nu}_{2p2h,11}$, describing processes in which the momentum ${\bf q}$ is transferred 
to a single high-momentum nucleon, requires the continuum component of the hole spectral function.

The second term in the right hand side of Eq.~\eqref{decompose}, involving the matrix elements of the two-nucleon current,  is written in terms of the two-nucleon spectral function \cite{spec2}.
The explicit expressions of $W^{\mu\nu}_{2p2h,11}$ and  $W^{\mu\nu}_{2p2h,22}$ are reported in Ref.~\cite{BLR}.

Finally,  $W^{\mu\nu}_{2p2h,12}$, taking into account interference contributions, involves the nuclear overlaps defined in both Eqs.~\eqref{def:phi2h1p} and \eqref{2hov0}. 
The resulting expression is

 \begin{widetext}
\begin{align} 
\label{total}
 W^{\mu\nu}&_{2p2h,\rm{12}} = \int d^3 k\ d^3\xi\  d^3\xi^\prime\ d^3h\  d^3 h^\prime d^3p\ d^3p^\prime 
{\Phi^{h h^\prime}_{\xi \xi^\prime}}^\ast  \left[ 
 {\Phi^{h h^\prime p^\prime}_{k}}  \langle {\bf k}|j^\mu_1|{\bf p}\rangle + {\Phi^{h h^\prime p}_{k}}  \langle {\bf k}|j^\mu_2|{\bf p^\prime}\rangle  \right] \\
 \nonumber
&  \times \langle {\bf p}{\bf p^\prime}| j^\nu_{12}| {\bm \xi}, {\bm \xi^\prime}\rangle  \  \delta({\bf h}+{\bf h^\prime}+ {\bf q} - {\bf p} - {\bf p^\prime})
\delta(\omega +e_{h}+ e_{h^\prime}-e_{p} - e_{p^\prime})
 \theta(|{\bf p}| - k_F)\theta(|{\bf p^\prime}| - k_F) + \rm{h.c.} \ .
\end{align}
\end{widetext}

We have compared the results of our approach to the measured electron-carbon cross sections in two different kinematical setups, 
corresponding to momentum transfer $300 \lesssim |{\bf q}| \lesssim 800\ {\rm MeV}$. The calculations have been carried out using the 
 carbon spectral function of Ref.~\cite{LDA} and the 1h contribution to the nuclear matter spectral function of Ref.~\cite{PKE}, as discussed in Ref.~\cite{BLR}.
The 2h1p amplitude, needed to evaluate the interference term, has been also computed  for nuclear matter at equilibrium density.
In the quasi elastic channel we have used the parametrization of the vector form factors of Ref. \cite{BBBA}, whereas the inelastic nucleon structure functions
have been taken from Refs.~\cite{BR1,BR2}. 

\begin{figure}[h!]
\begin{center}
\includegraphics[scale=0.6]{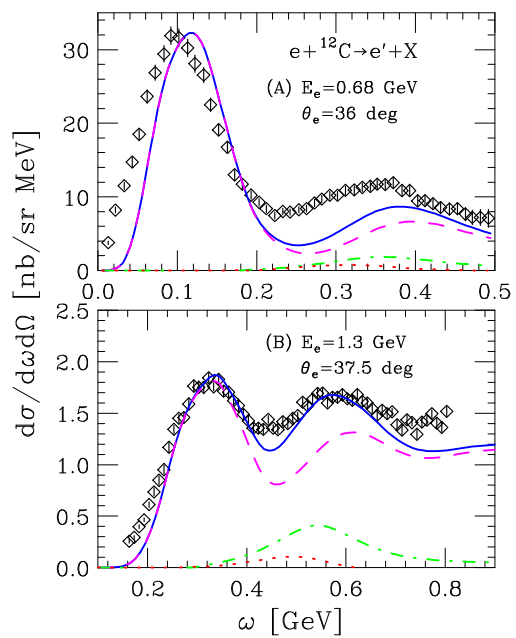}
\end{center}
\vspace*{-.2in}
\caption{(color online) (A): Double differential cross section of the process $e+{^{12}C}\to e^\prime +X$ at beam energy $E_e=680 $ MeV and scattering angle $\theta_e= 37.5$ deg. The solid line shows the 
result of the full calculation, while the dashed line has been obtained including the one-body current only. 
The contributions arising from the two-nucleon current are illustrated by the dot-dash and dotted lines, corresponding to the pure two-body current transition probability and 
to the interference term, respectively. The experimental data are taken from Ref.~\cite{barreau}. (B) same as (A) but for $E_e= 1300$ MeV and $\theta_e= 37.5$ deg.The experimental data are taken from Ref.~\cite{carbon:data}.}
\label{xsec:680:1300}
\end{figure}

Figure~\ref{xsec:680:1300} shows the electron-carbon cross section 
at beam energy $E_e=680$~MeV and scattering angle $\theta_e= 36$~deg (A) , $E_e= 1300$~MeV and $\theta_e= 37.5$~deg (B) . The solid and dashed lines correspond to the results of the full calculation 
and to the one-body current contribution, respectively. The pure two-body current contribution and the one arising from interference are illustrated by the dot-dash and dotted line, respectively. 
In the kinematics of panel (A) the two-body currents play an almost negligible role. The significant lack of strength in the $\Delta$-production region,  discussed in Ref.~\cite{BM}, 
is likely to be due to inadequacy of the structure functions of Refs.~\cite{BR1,BR2} to describe the region of $Q^2 \lsim 0.2$ GeV$^2$, while the shift in the position of the quasi-elastic peak has to be ascribed to the effects of FSI, which are not taken into account. 

At the larger beam energy and $Q^2$ corresponding to panel (B), the agreement between theory and data is significantly improved, and the 
contribution of the two-nucleon current turns out to substantially increase the cross section in the dip region.

In inclusive processes, FSI have two effects:  a shift of the cross section, arising from the interaction between the struck nucleon and the mean field generated by the spectator particles, and 
a redistribution of the strength from the quasi-elastic peak to the tails. The theoretical approach for the description of  FSI within the spectral function formalism is discussed in 
Refs.~\cite{RMP,PRD,benharFSI}. 

According to Ref.~\cite{benharFSI}, the differential cross section can be  written in the convolution form
\begin{align}
d\sigma_{FSI}(\omega)= \int d\omega^\prime f_{\bf q}(\omega - \omega^\prime - U_V)d\sigma(\omega^\prime)\ ,
\label{conv:sigma}
\end{align}
where $d\sigma$ denotes the cross section in the absence of FSI, the effects of which are accounted for by the  
folding function
\begin{align}
f_{\bf q}(\omega)=
 \sqrt{T_A} \delta(\omega)+(1-\sqrt{T_A})F_{\bf q}(\omega)\ .
\label{folding:func}
\end{align}
The above equations show that FSI are described in terms of the real part of the optical potential $U_V$, extracted from proton-carbon scattering data~\cite{optpot}
responsible for the shift in $\omega$,  the nuclear transparency $T_A$, measured in coincidence $(e,e^\prime p)$ reactions~\cite{transparency}, and
a function $F_{\bf q}(\omega)$, sharply peaked at $\omega = 0$, whose width is dictated by the NN scattering cross section~\cite{benharFSI}. 

A comprehensive analysis of FSI effects on the electron-carbon cross sections has been recently carried out by the authors of  Ref.~\cite{uncertainty}. 
In this work we have followed closely their approach, using the same input.

\begin{figure}[h!]
\begin{center}
\includegraphics[scale=0.6]{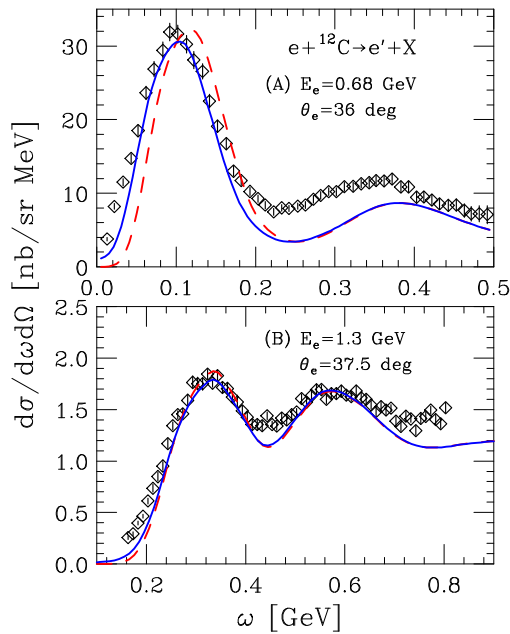}
\end{center}
\vspace*{-.2in}
\caption{(color online) (A): double differential electron-carbon cross section at beam energy $E_e=680 $ MeV and scattering angle $\theta_e= 36$ deg. The dashed line corresponds to the result obtained neglecting FSI, while the solid line has been obtained within the approach of Ref.~\cite{uncertainty}.  The experimental data are taken from Ref.~\cite{barreau}. (B): same as (A) but for $E_e= 1300$ MeV and $\theta_e= 37.5$ deg. The experimental data are taken from Ref.~\cite{carbon:data}.
 }
\label{680_1300_fsi}
\end{figure}

Figure~\ref{680_1300_fsi} illustrates the effects  of FSI on the electron-carbon cross section in the same kinematical setups of Fig.~\ref{xsec:680:1300}. In panel (A), both the pronounced shift of the 
quasi elastic-peak, and the redistribution of the strength are clearly visible, and significantly improve the agreement between theory and data. 
 For larger values of $Q^2$,  however, FSI play a less relevant role. This feature is illustrated in panel (B),  showing that 
 at beam energy $E_e=1.3$ GeV and scattering angle $\theta_e=37.5$ deg, corresponding to $Q^2 \sim 0.5$ GeV$^2$, the results of  calculations carried out with and without inclusion of FSI
 give very similar results, yielding a good description of the data.
 
Note that, being tranverse in nature, the calculated two-nucleon current contributions to the cross sections exhibit a strong angular dependence.
At $E_e=1.3$~GeV, we find that the ratio between the integrated strengths in the 1p1h and 2p2h sectors grows from  4\% at electron scattering angle $\theta_e$=10 deg
to 46\% at $\theta_e$=60 deg.


The results of our work show that the approach based on the generalized factorization {\em ansatz} and the spectral function formalism
provides a consistent framework for a unified description of the electron-nucleus cross section, in the kinematical regime in which relativistic effects are known to be important.   

The extension of our approach to neutrino-nucleus scattering, which does not involve additional conceptual difficulties, will 
offer new insight on the interpretation of the cross section measured by the MiniBooNE Collaboration in the quasi elastic channel \cite{CCQE,NC}. 
The excess strength in the region of the quasi elastic peak is in fact believed to originate from processes involving two-nucleon currents  \cite{nieves,martini,superscaling}, whose contributions is observed at  lower energy loss as a result of the average over the neutrino flux \cite{nufact_ginevra}.  
The strong angular dependence of  the two-nucleon current contribution, may also provide a clue for the understanding of the differences between the quasi elastic cross sections reported by the MiniBooNE and 
NOMAD Collaboration \cite{nomad}, which collected data using neutrino fluxes of mean energies 880 MeV and 25~GeV, respectively \cite{nufact_ginevra}.

As a final remark, it has to be pointed out that a clear-cut identification of the variety of reaction mechanisms contributing to the neutrino-nucleus cross section will require a careful analysis 
of the dynamical assumptions underlying the different models of nuclear dynamics. All approaches 
based on the independent particle model of the nucleus fail to properly take into account correlation effects, leading to a significant reduction of the normalization of  
the shell-model states\textemdash unambiguously observed in $(e,e^\prime p)$ experiments \cite{eep}\textemdash as well as to the appearance of sizable interference terms in the 2p2h sector.
However, in some instances these two deficiencies may largely compensate one another, leading to {\em accidental} agreement between theory and data. For example, the two-body current 
contributions computed within our approach turn out to be close to those obtained from the Fermi gas model. The development of a nuclear model having the predictive power needed for 
applications to the analysis of future experiments\textemdash most notably the 
Deep Underground Neutrino Experiment (DUNE)~\cite{DUNE}\textemdash will require that the degeneracy between different approaches be resolved. A systematic comparison between the 
results of theoretical calculations and the large body of electron scattering data, including both inclusive and exclusive cross sections, will greatly help to achieve this goal.

This research is supported by INFN (Italy) under grant MANYBODY (NR and OB)  and the U.S. Department of Energy, Office of Science, Office of
Nuclear Physics, under contract DE-AC02-06CH11357 (AL).

\end{document}